# Harmonizing 3GPP and NFV description models to provide customized RAN slices in 5G networks

Oscar Adamuz-Hinojosa[1,2], Pablo Munoz,[1,2] Jose Ordonez-Lucena[3], Juan J. Ramos-Munoz,[1,2] and Juan M. Lopez-Soler[1,2]

[1] Research Center on Information and Communication Technologies, University of Granada.

[2] Department of Signal Theory, Telematics, and Communications, University of Granada.

[3] Technology Exploration & Standards, GCTIO Unit, Telefonica I+D.

*The standardization of Radio Access Network (RAN) in mobile networks has traditionally been led by 3GPP. However, the emergence of RAN slicing has introduced new aspects that fall outside 3GPP scope. Among them, network virtualization enables the particularization of multiple RAN behaviors over a common physical infrastructure. Using Virtualized Network Functions (VNFs) that comprise customized radio functionalities, each virtualized RAN, denominated RAN slice, could meet its specific requirements. Although 3GPP specifies the description model to manage RAN slices, it can neither particularize the behavior of a RAN slice nor leverage the NFV descriptors to define how its VNFs can accommodate its spatial and temporal traffic demands. In this article, we propose a description model that harmonizes 3GPP and ETSI-NFV viewpoints to manage RAN slices. The proposed model enables the translation of RAN slice requirements into customized virtualized radio functionalities defined through NFV descriptors. To clarify this proposal, we provide an example where three RAN slices with disruptive requirements are described following our solution.*

## Introduction

The fifth generation (5G) networks aim to boost the digital transformation of industry verticals. These verticals may bring a wide variety of unprecedented services with diverging requirements in terms of functionality and performance. Considering each service separately and building a Radio Access Network (RAN) accordingly would be unfeasible in terms of cost. To economically provide these services, RAN slicing has emerged as a solution [1]. It consists of the provision of multiple RAN slice subnets, each adapted to the requirements of a specific service, over a common wireless network infrastructure.

The leading standardization body on RAN slicing is the Third Generation Partnership Project (3GPP). It defines a RAN slice subnet as a set of next Generation NodeBs (gNBs) that are arranged and configured to provide a particular RAN behavior. To manage its lifecycle, the 3GPP defines the RAN Network Slice Subnet Management Function (NSSMF) and the Network Function Management Functions (NFMFs) as the management entities; and the RAN Network Slice Subnet Template (NSST) as the deployment template [2].

To achieve the flexibility and modularity that a RAN slice subnet requires, some gNB functionalities can be implemented by software, i.e., by Virtualized Network Functions (VNFs) [3]. However, the lifecycle management of VNFs and the orchestration of their resources goes beyond 3GPP scope. The European Telecommunication Standard Institute (ETSI), specifically the Network Function Virtualization (NFV) group, is playing a significant role on these tasks. To that end, ETSI-NFV has defined the NFV Management and Orchestration (MANO) and NFV descriptors.

Focusing on RAN slicing descriptors, the RAN NSST considers the gNB functionalities of a RAN slice subnet. However, the 3GPP has not specified how these functionalities must be configured to meet the requirements for a specific service, typically enhanced Mobile Broadband (eMBB), ultra-Reliable Low Latency Communication (uRLLC), and massive Machine Type Communication (mMTC). Additionally, the RAN NSST neglects the resource requirements for the virtualized deployment of some gNB functionalities. For this, the RAN NSST could use the NFV descriptors. Notwithstanding, describing the virtual resources to accommodate the fluctuations of spatial and temporal traffic demands of a RAN slice subnet is a challenge.

Recent works have addressed the description of RAN slice subnets. For instance, the authors of [4] propose a set of configuration descriptors to parametrize the features, policies and radio resources within the gNBs of a RAN slice subnet. With these descriptors, this work provides a first attempt to define the customized behavior of a RAN slice subnet. However, 3GPP completed the New Radio (NR) specifications after that work, thus the impact of the NR parameters in RAN have not been analyzed in depth yet. Additionally, although this work considers partially-virtualized gNBs, it neglects the description of the virtual resources required to build up them. Thereby, describing the spatial and temporal traffic demands of a RAN slice subnet with NFV descriptors is still an open question.

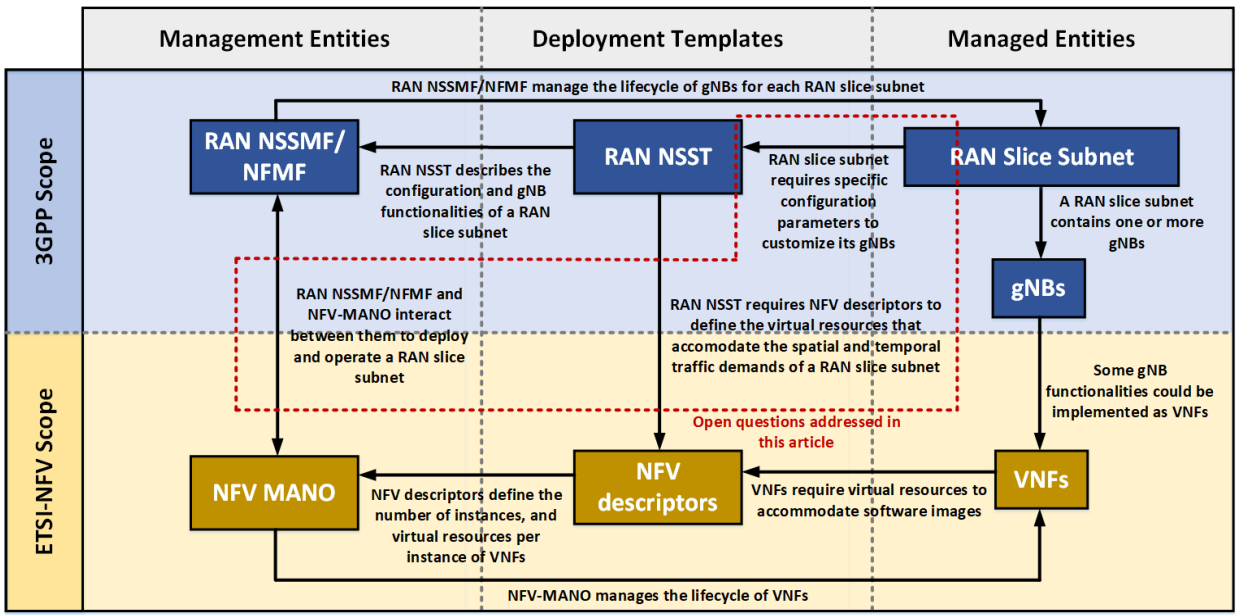

*Figure 1* Relationship between the 3GPP and ETSI-NFV scopes for the deployment and operation of RAN slices subnets. The aspects within the dotted box are open questions that are addressed in this article.

In this article, we provide a description model for RAN slicing. By harmonizing 3GPP and ETSI-NFV scopes, the proposed solution allows the management of virtualized gNB functionalities, and their customization by setting predefined radio parameters. Thereby, an operator could efficiently provide RAN slice subnets to accommodate the services demanded by verticals on a geographical area with specific spatial and temporal traffic demands. To gain insight into this proposal, we provide an example where RAN slice subnets for eMBB, uRLLC, and mMTC are described based on the proposed solution. For comprehensibility purposes, Fig. 1 illustrates the context and the addressed issues of this article.

## RAN slicing enablers

### NG-RAN architecture

The 3GPP has defined the Next Generation RAN (NG-RAN) as the 5G RAN architecture. This architecture comprises gNBs connected to the 5G Core Network. Each gNB provides NR user/control plane protocol terminations towards the User Equipments (UEs). In turn, each gNB comprises one Centralized Unit (CU), multiple Distributed Units (DUs) and multiple Radio Units (RUs) [5].

As depicted in Fig. 2, the gNB functionalities are distributed over CU, DUs and RUs in a flexible way. The RUs comprises at least radio-frequency circuitry, thus their functionalities are implemented as Physical Network Functions (PNFs), i.e., dedicated hardware appliances. The remaining functionalities, gathered in the DUs and the CU, may be virtualized as VNFs. The DUs contain low-layer functionalities whereas the CU includes high-layers functionalities. According to 3GPP, there exists up to eight options to split radio functionalities between the CU and DUs. The aim of functional split is to leverage the benefits of virtualization (e.g., reducing costs and dynamic scalability) and centralization (e.g., statistical multiplexing gains).

However, the majority of these options present a set of issues and challenges that will difficult their short-term implementation [5]. For this reason, there is a consensus in the industry and academia that the most feasible implementation is the option #2 for splitting CU-DUs. This option could be implemented on the basis of Dual Connectivity (DC) standard.

Regarding the functional split for DUs-RUs, the Common Public Radio Interface (CPRI) has arisen as a standard for implementing option #8. It enables the transmission of baseband signals over transport links. The main drawback of this option is the higher capacity required for these links. To relieve the data rate demands between DUs and RUs, the evolved CPRI (eCPRI) standards proposed aggregating the low-layer functionalities of PHY in the RU, resulting in the split option #7. Furthermore, eCPRI allows an efficient and flexible radio data transmission via a packet-transport network like IP or Ethernet. However, the aggregation of Low-PHY functionalities leads to a significantly higher cost of RUs. In this article, we assume that the implementation of split options #7 or #8 will depend on the features of the transport network in each deployment area.

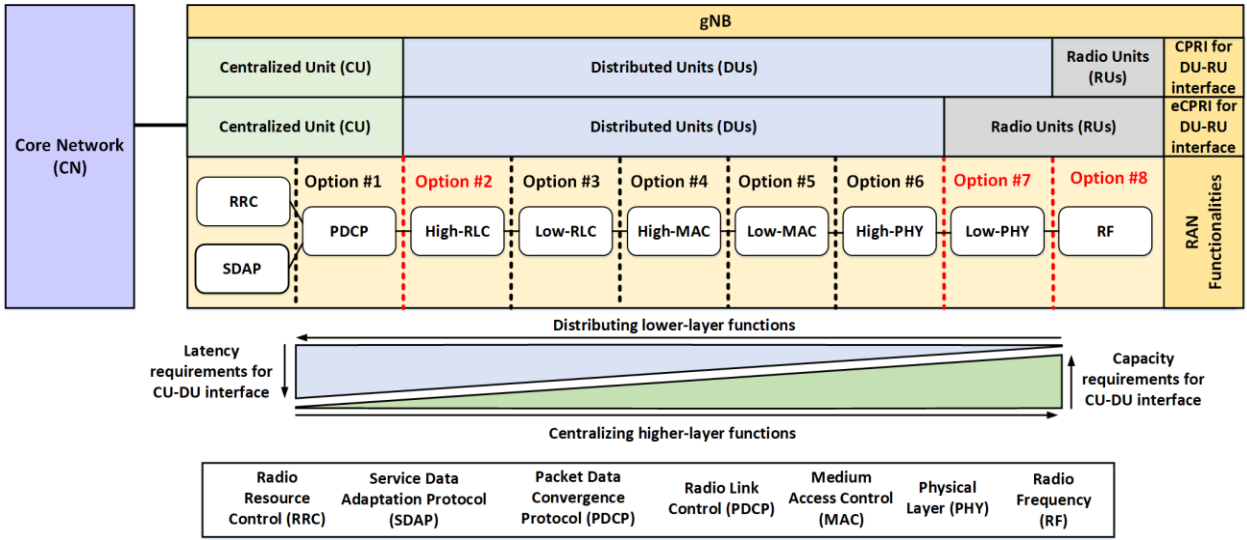

***Figure 2*** *3GPP functional split options for the gNB. Among these split options, the option #2 is the best candidate for CU-DUs splitting and the options #7 and #8 for DUs-RUs splitting in short-term deployments. Note that the latency requirements for CU-DU interface refers to the maximum tolerable latency provided by this transport link. Above this value, the data transmission between CU and DU would be desynchronized.*

In short-term deployments, the CU will be executed as a VNF in a Point of Presence (PoP), i.e., a cloud site where VNFs can run, while DUs will be likely implemented as PNFs. There are two main reasons. First, the software images of DUs must be optimized to execute ms procedures. Secondly, to satisfy the stringent latency requirements, PoPs hosting DUs must be installed near users, even closer the PoPs hosting CUs.

Despite these issues, researchers are working on the DU virtualization. Some works (e.g., [6]) consider a hierarchical structure of PoPs to enable the virtualization of both, the CU and the DU. Furthermore, some gNB software implementations (e.g., OpenAirInterface [6]) are considering the CU-DUs split.

Assuming virtualized CU and DUs in this article, the RAN infrastructure requires a hierarchical structure of PoPs in addition to cell sites, as depicted in Fig. 3. These PoPs might be hosted in the aggregation and distribution nodes that connect the cell sites with the Core Network [7]. Since an aggregation node serves multiple RUs, the hosted PoP, could allocate DUs per each RAN slice subnet that requires the coverage area of these RUs. Similarly, the PoP hosted in a distribution node could allocate CUs serving the DUs of each RAN slice subnet.

Focusing on an aggregation PoP, if the geographical region served by this PoP has a high UE density, the allocated DU of a RAN slice subnet will usually serve more cell sites, thus requiring more virtual resources to deal with the aggregated traffic. Similarly, a DU serving a region with low cell sites density, will usually require less virtual resources.

In an edge PoP, the amount of virtual resources required by a CU depends on the number of served DUs and the cell sites density supported by each DU.

*3GPP RAN slicing management functions and descriptor*

To manage the lifecycle of RAN slice subnets, the 3GPP has defined the RAN NSSMF and the NFMFs [2]. The RAN NSSMF (a) translates the performance and functional requirements of a gNB into the amount of the virtual resources that accommodate the gNBs; and (b) manages the Fault, Configuration, Accounting, Performance, and Security (FCAPS) of the gNBs from the application perspective. Each NFMF is specific for a type of gNB component (i.e., CU, DUs, or RUs), and is controlled by the RAN NSSMF to carry out the activities related to (b).

To automate the lifecycle management of each RAN slice subnet, the RAN NSSMF uses RAN NSSTs. Each RAN NSST defines the gNB functionalities, and their specific configuration to meet the specific performance requirements of a service type (i.e., eMBB, uRLLC, and mMTC). To identify this service type, the RAN NSST contains the Single Network Slice Selection Assistance Information (S-NSSAI) [8]. This 3GPP parameter consists of two fields: Slice/Service Type (SST) and Slice Differentiator (SD). SST provides a value that identifies the service type of the slice, i.e., SST=1 for eMBB, SST=2 for uRLLC, and SST=3 for mMTC. SD is optional and allows differentiation amongst multiple network slices with the same SST value, e.g., slices for different tenants.

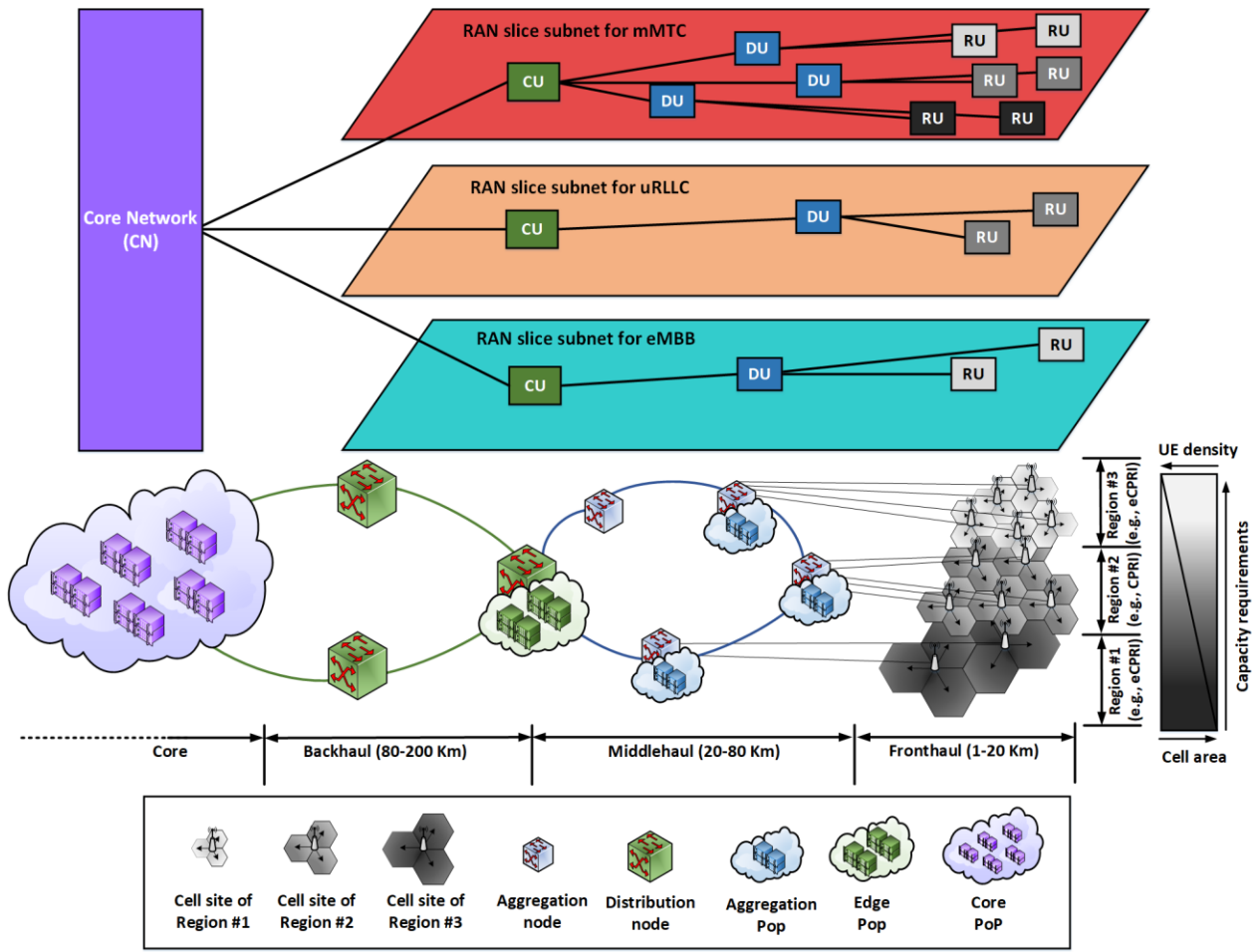

**Figure 3** Deployment perspective of RAN slice subnets for mMTC, uRLLC, and eMBB, respectively. By way of example, the RAN slice subnet for mMTC is deployed over the three regions. The RAN slice subnet for uRLLC is deployed over the Region #2. The RAN slice subnet for eMBB is deployed over the Region #1. Furthermore, fronthaul links for Regions #1 and #3 use eCPRI whereas for Region #2 use CPRI.

### NFV MANO and descriptors

To manage VNFs, ETSI-NFV has defined the NFV MANO [9]. It comprises:
- Virtualized Infrastructure Manager (VIM), which manages the virtual resources from one or PoPs.
- VNF Manager (VNFM), which manages the VNFs throughout their lifecycle. It is also responsible for their performance and fault management from the virtualization viewpoint.
- NFV Orchestrator (NFVO), which combines PNFs and VNFs to create network services, managing them throughout their lifecycle.

To automate the lifecycle management of network services and their VNFs and/or PNFs, the NFV-MANO uses the NFV descriptors: Network Service Descriptor (NSD), VNF descriptor (VNFD) and PNF Descriptor (PNFD).

Each NSD (and VNFD) defines a set of attributes. Among them, the flavors provide different options to deploy an instance of a network service (and VNF). For example, each flavor might add some extra functionalities to that instance. In turn, each flavor defines one or more instantiation levels (ILs), each specifying a different amount of virtual resources for the instance deployed from that flavor. Defining several ILs enables the adaptation of the required amount of virtual resources to guarantee the performance of an instance of network service (and VNF) supporting traffic fluctuations. For more detailed information about flavors and ILs, see [9].

Finally, since NFV-MANO focuses on virtualization, the PNFDs only contain information required to connect PNFs with VNFs.

| Configuration parameters to customize the behavior of a RAN slice subnet | | RAN slice subnet requirements | | | | | |
|---|---|---|---|---|---|---|---|
| | | Latency | Maximum Mobility speed | Throughput per UE | UE density | Reliability | Priority Level |
| 3GPP NR | Waveform and numerology | ✓ | ✓ | | | ✓ | |
| | Operation bands | ✓ | ✓ | ✓ | ✓ | ✓ | |
| | Slot Format | | | ✓ | ✓ | | |
| | 5G QoS Indicators (5QIs) | ✓ | | | | ✓ | ✓ |
| | Modulation and Coding Schemes (MCS) | | | ✓ | | | |
| Network management algorithms | e.g., Radio Resource Management (RRM); and Self-Organizing Network (SON) techniques | ✓ | | ✓ | ✓ | ✓ | ✓ |

*Table 1* Relationship between RAN slice subnet requirements and the configuration parameters to customize the behavior of the gNB functionalities for a RAN slice subnet.

## RAN slice description proposal

*Harmonizing 3GPP and NFV descriptors: A prerequisite for managing RAN slice subnets*

To manage the gNBs taking part in each RAN slice subnet, the RAN NSSMF must rely on RAN NSSTs and NFV descriptors.

On the one hand, the RAN NSST focuses on the description of the gNBs of a RAN slice subnet from an application perspective (i.e., information on their functionalities and configuration parameters). The aim of a RAN NSST is to adapt the behavior of the gNBs to meet the requirements of a specific service type (e.g., eMBB). However, the RAN NSST neglects the description of the resources to deploy the virtualized part of these gNBs.

On the other hand, the NFV provides information on the virtual resources that are required to accommodate the spatial and temporal traffic demands of the CU and DUs of a gNB. This means that NFV descriptors could enable the deployment of the virtualized part of a gNB. However, NFV descriptors are agnostic to the application layer configuration of the CU and DUs.

With the combined use of 3GPP and NFV descriptors, the gNBs of a RAN slice subnet could be deployed and operated. Accordingly, we first analyze the most representative configuration parameters to customize the behavior of a gNB. Then, we propose a description model that harmonizes the scopes of the RAN NSSTs and NFV descriptors to manage the gNBs taking part in different RAN slice subnets. Finally, we explain how the RAN NSSMF and NFMFs interwork with the NFV-MANO to manage RAN slice subnets with the proposed model and configuration parameters.

*Configuration parameters in a RAN NSST*

According to Table 1, the most representative parameters are classified into two groups: 3GPP NR, and network management algorithms.

The 3GPP NR comprises those parameters related to the physical transmission. Among them, the waveform and numerology, the operations bands, the slot format, the 5G QoS Indicators (5QIs), and the Modulation and Coding Schemes (MCSs) are discussed below.

The waveform is based on Orthogonal Frequency Division Multiplexing (OFDM). It consists of several orthogonally-spaced subcarrier with a spacing of $15 \cdot 2^\mu$ KHz [10], where $\mu$ is the numerology ($\mu$=0, 1, 2, 3 and 4). The higher the numerology is, the shorter the Transmission Time Interval (TTI) is. Decreasing the TTI enables gNBs to transmit UE data faster; and add a margin to increase the number of retransmissions in the hybrid automatic repeat request function. Therefore, shorter TTIs are suitable for RAN slice subnets that require low latency and high reliability. Additionally, high-speed UEs can benefit from shorter TTIs, taking advantage of the time invariant characteristics of the channel.

The NR operation bands includes 450-6000 MHz and 24250-52600 MHz [11]. Each band might accommodate carriers from 5 to 400 MHz. The bandwidth of the selected carrier depends on the required service data rate and the UE density in the geographical regions where the RAN slice subnet is deployed.

The selection of the operation bands also fixes the transmission mode, i.e., Frequency Division Duplex (FDD) or Time Division Duplex (TDD). In case of the TDD mode, there exists predefined slot formats that assign downlink and uplink bits at OFDM symbol level [10]. The selection of the slot format for a given RAN slice subnet depends on the symmetry between its downlink and uplink requirements.

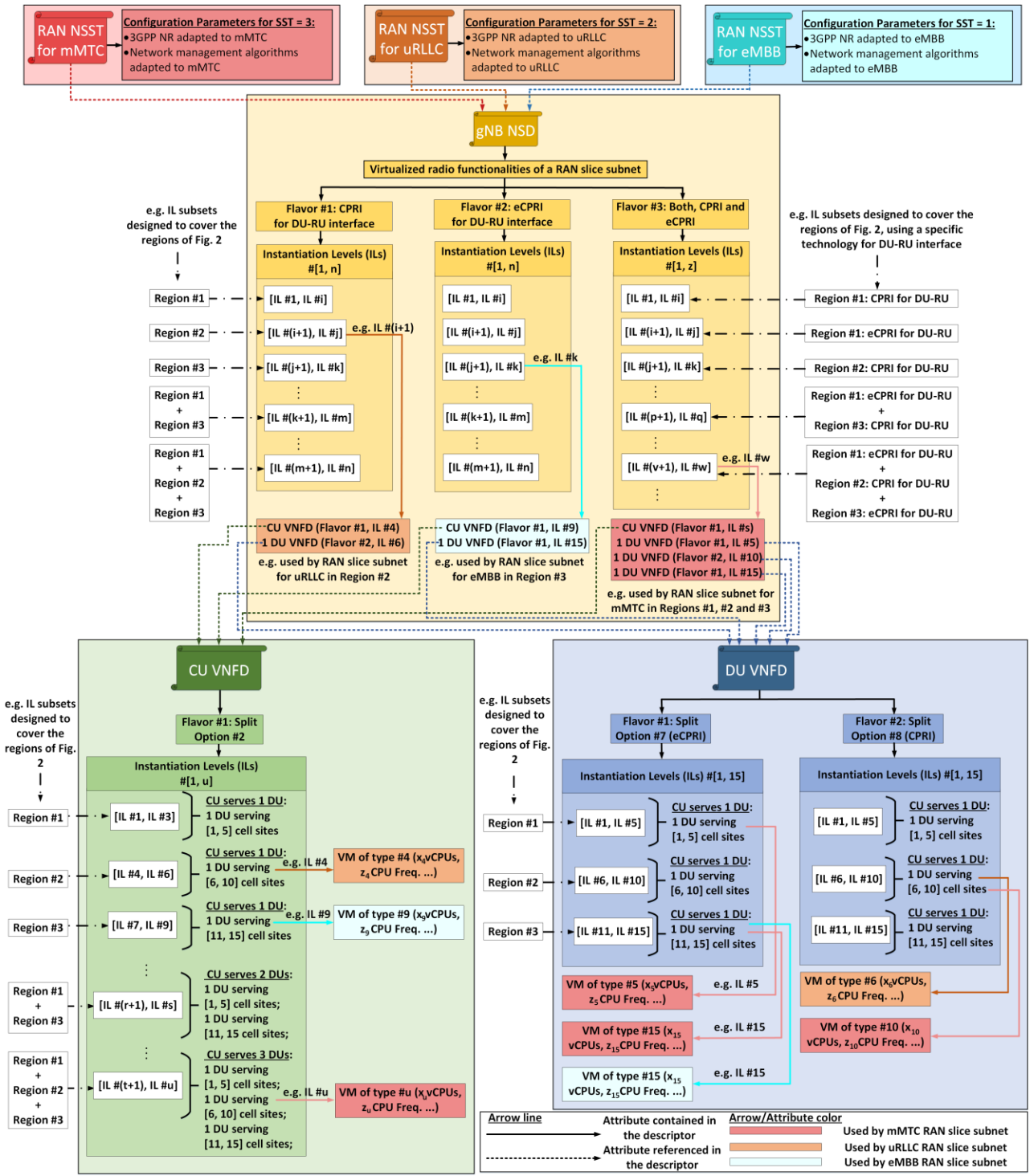

**Figure 4** Proposed model to define the management of a gNB for each RAN slice subnet. By way of example, the gNBs of the three RAN slice subnets presented in Fig. 3 are described. To deploy these gNBs, the RAN NSSMF selects in the gNB NSD the tuples (Flavor #3, IL #w), (Flavor #1, IL #(i+1)) and (Flavor #2, IL #k) for mMTC, uRLLC and eMBB RAN slice subnets, respectively. Note that the mMTC RAN slice subnet requires both, the CPRI and eCPRI for DU-RU interfaces.

The 5QI specifies the class that ensures a specific QoS forwarding behavior in the RAN domain [8]. Each class is mainly characterized by a priority level, a packet delay budget and a packet error rate. Each parameter has a direct impact on the performance of a RAN slice subnet. For example, with the packet delay budget and the packet error rate, the RAN NSSMF can control the latency and reliability level of the RAN slice subnet, respectively. Similarly, with the priority level the RAN NSSMF can weigh the utilization of radio resources shared among the RAN slice subnets, providing in this way multiplexing gains.

The MCS is a modulation scheme and coding rate tuple that provides a given throughput for an UE. Each gNB selects the MCS per each UE based on its current radio conditions. NR defines two set of MCS tuples: one is compatible with the MCSs defined in LTE and the other extends that range to include a higher modulation scheme, thus enabling higher throughput in NR. Each RAN slice subnet should only use the set of MCSs that best meet its throughput requirements.

Network management algorithms are usually proprietary and include vendor-specific parameters. However, some parameters could be configured by the RAN NSSMF, allowing the definition of slice-specific network management algorithms that optimize the operation of each RAN slice subnet. Network management includes traditional Radio Resource Management (RRM) functionalities (e.g., packet scheduling) and Self-Organizing Network (SON) techniques (e.g., mobility robustness optimization).

*Description model to manage RAN slice subnets*

Fig. 4 shows the proposed description model to define the management of gNBs of several RAN slice subnets. Each RAN NSST references (a) a common NSD that describes the underlying resources of a gNB; and (b) contains the specific configuration parameters for this gNB (i.e., those adapted to a specific SST).

The gNBs of any RAN slice subnet may be stick to the option #2 for CU-DU functional split. However, the option selected for DU-RU split can vary between #7 and #8, depending on the technology used for the underlying fronthaul links, i.e., CPRI or eCPRI. For this reason, the gNB NSD defines three flavors: one supporting only CPRI, other supporting only eCPRI, and the other supporting the joint usage of both technologies in case that they were implemented in a specific deployment area.

Each flavor in the gNB NSD defines different subsets of ILs depending on the type of region(s) (e.g., those represented in Fig. 3) covered by a gNB. Since the number of cell sites located in each region is different, each subset gathers those ILs adapted to the possible range of the aggregated traffic demands for a certain number of cell sites located in one or more regions. In the case of using flavor #3, the technology for implementing the DUs-RUs interface also conditions the IL subsets (e.g., two IL subsets for Region #1, one describing CPRI interfaces and the other describing eCPRI interfaces).

For a given subset, each IL defines the number of DUs in the gNB as well as the virtual resources required to deploy these DUs and the CU. The amount of virtual resources is completely different for the CU and DUs. Whereas each DU may serve a number of cell sites in a region, a CU may aggregate the traffic from DUs serving one or more regions, each with different cell sites density. For that reason, the gNB NSD distinguishes between the resources requirements of CU and DU by referencing a different VNFD (with specific flavor and IL) in each case.

Focusing on DU VNFD, it contains two flavors for the specification of the DU-RU functional split. Each flavor enables the split option #7 and #8, respectively. In turn, each flavor defines subsets of ILs, each gathering those ILs adapted to the possible traffic demands from a specific range of number of cell sites served by a DU. Each IL defines the characteristics (i.e., number of cores, CPU frequency, RAM capacities, etc.) of the Virtual Machine (VM) that hosts the DU functionalities. The utilization of this VM mainly depends on aspects such as the amount of radio resources and the MCSs used per each UE [12].

Similarly, the CU VNFD contains one flavor to define the split option #2 for CU-DUs. Each flavor also contains subsets of ILs. However, in this case, these subsets define ILs to support a specific number of DUs since a CU might aggregate DUs from different regions. Depending on the number of DUs served by the CU and their capacities, the characteristic of the VM that host the CU differs between ILs.

Finally, since RUs are fixed in specific locations, the gNB NSD cannot include references to the PNFDs to be reusable in any deployment area. The RAN NSSMF is responsible for selecting the specific PNFDs to define the RUs of a RAN slice subnet.

*RAN NSSMF, NFMFs and NFV-MANO interworking under a unified framework*

To manage the gNBs taking part in a RAN slice subnet, there is a necessity to define a unified framework where 3GPP entities (i.e., RAN NSSMF, NFMFs) and ETSI NFV (NFV-MANO) can work with each other. Examples of tentative integration have been proposed in [2-3]. In Fig. 5, this unified framework is depicted. Each management entity is common for all RAN slice subnets.

When a vertical requests a service for a specific geographical area, the RAN NSSMF first selects a RAN NSST whose SST matches with the requested service. Then, the RAN NSSMF determines which RUs cover the geographical area.

Once the RUs are selected, the RAN NSSMF computes the number of gNBs that will include these RUs. To that end, based on the UE density in each region of the deployment area, the RAN NSSMF performs the following actions:
- Select the flavor of the gNB NSD according to the technology of fronthaul links (i.e., flavor #1 for CPRI or flavor #2 for eCPRI). If this deployment area comprises fronthaul networks using both technologies, the flavor #3 is selected, since it enables the definition of a DU-RU interface with any of these technologies. For example, the mMTC RAN slice subnet shown in Fig. 3 requires the flavor #3 because Regions #1 and #3 usee eCPRI fronthaul links, and Region #2 uses CPRI fronthaul links.
- Compute the number of DUs. To that end, for every region on the deployment area, the RAN NSSMF determines the optimal IL subset for one DU. This IL subset can accommodate the fluctuations of the temporal traffic demands of this area. If this DU cannot serve the entire region, additional DUs with a specific IL subset are included to meet the required capacity in this region. Thereby, the spatial traffic demands of this area are also accommodated.
- Determine the number of CUs that will serve the DUs. Considering the latency constraints due to the physical distance between a CU and the selected DUs, the RAN NSSMF (a) optimally distribute these DUs between a specific number of CUs; and (b) select the optimal IL subset for each CU. The aim of this procedure is minimizing the number of CUs to benefit from the statistical multiplexing gains provided by this centralization approach. The number of CUs is equivalent to the number of gNBs.
- Search across the IL subsets of the gNB NSD, the subset that reference the selected IL subset for each DU and the CU that serves these DUs. Thereby, the RAN NSSMF derives the optimal IL subset for each gNB.

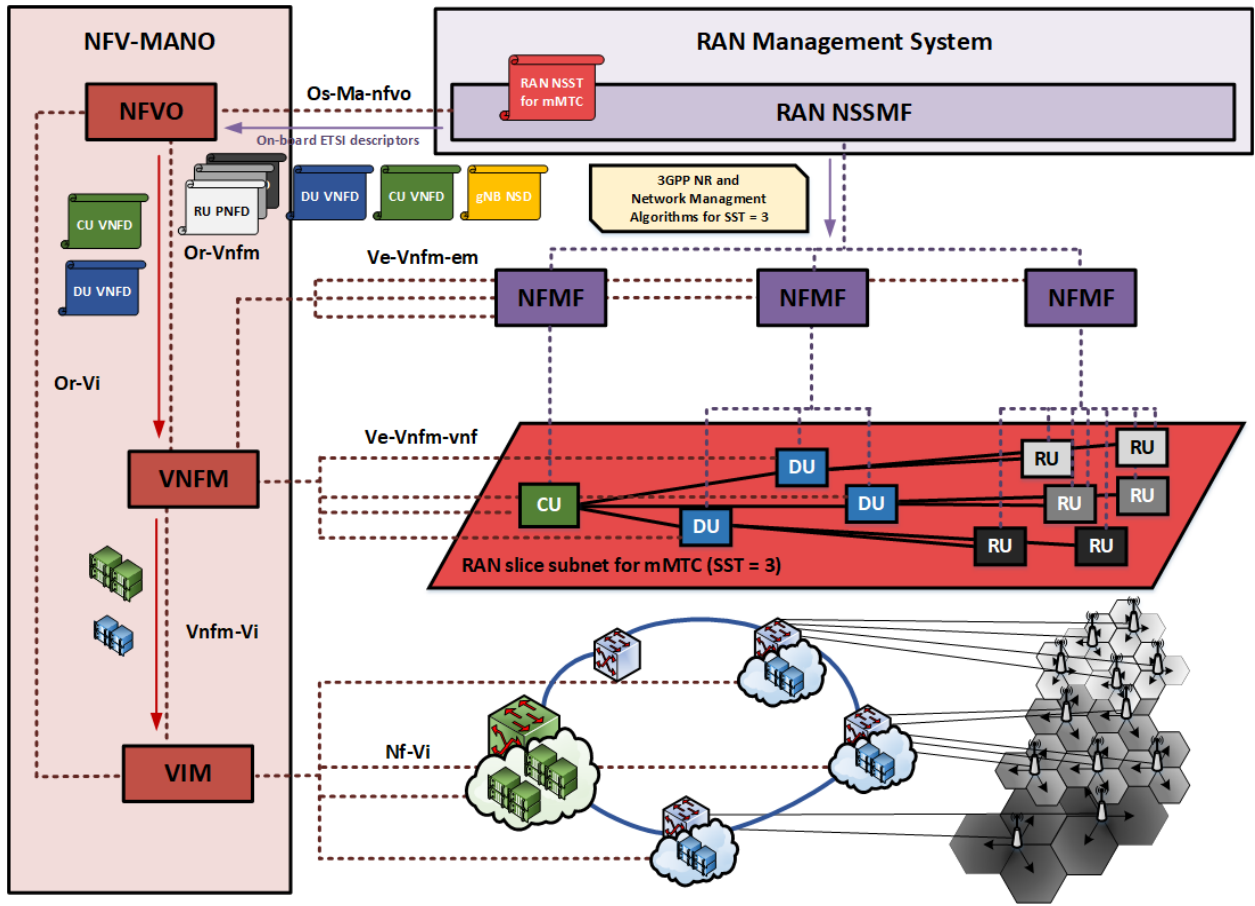

**Figure 5** RAN slicing management framework. By way of example, this framework manages the deployment and operation of the RAN slice subnet for mMTC (see Fig. 3 and 4).

Next, the RAN NSSMF proceeds with the on-boarding of the NFV descriptors along with the selected flavor and the IL subset per each gNB. With this information, the NFVO can instantiate the gNBs and scale them throughout the lifecycle of the RAN slice subnet. The VNFM and VIM also play a key role during the lifecycle through the management of CU/DUs and their underlying virtual resources, respectively. For more information, see [9].

To customize the behavior of the gNBs, the RAN NSSMF uses the configuration parameters defined in the RAN NSST. With this information, the RAN NSSMF configures the CU and DUs through the specific NFMFs, which apply the parameters provided by the RAN NSSMF.

### Example of RAN slice description

To clarify our proposal, this section provides an example where RAN slice subnets for eMBB, uRLLC, and mMTC are deployed over the same city (see details in Table 2). These RAN slice subnets can be mapped to the ones presented in Fig. 3.

| RAN Slice Subnet Requirements | RAN Slice Subnets | | |
|---|---|---|---|
| | eMBB (e.g., UHD streaming) | mMTC (e.g., pollution control) | uRLLC (e.g., remote controlled drones) |
| Latency (ms) | 10 | Seconds to hours | 5 |
| Maximum Mobility Speed (Km/h) | 10 | 0 | 250 |
| Throughput per UE — Uplink (Mbps): | 50 | 0.1 | 25 |
| Throughput per UE — Downlink (Mbps): | 300 | 0.1 | 1 |
| UE density | 5.000 UEs/Km$^2$ | 500.000 UEs/Km$^2$ | 50 UEs/Km$^2$ |
| Reliability (%) | Not specified | Not specified | 99.999 |
| Priority Level | Low | Medium | High |
| UE type | Pedestrians | Stationary sensors | Remote-controlled vehicles |
| Geographical regions | City center | City center, industrial area, and suburban area | Suburban area |

**Table 2** RAN slice subnet requirements for eMBB [13], mMTC [13] and uRLLC [14]. Note that the geographical regions might be mapped to the ones presented in Fig. 3. More precisely, industrial area to Region #1, suburban area to Region #2 and city center to Region #3.

| Configuration Parameters | | RAN Slice Subnets | | |
|---|---|---|---|---|
| | | RAN NSST for eMBB | RAN NSST for mMTC | RAN NSST for uRLLC |
| 3GPP NR | Waveform and numerology | μ = 2 | μ = 0 | μ = 3   Note 1 |
| | Operation bands | 450-6000 MHz (max. carrier bandwidth 100 MHz), 24250 to 52600 MHz (max. carrier bandwidth 400 MHz) | 450-6000 MHz (carrier bandwidth 5 MHz) | 24250 to 52600 MHz |
| | Slot Format | #28 (12 OFDM symbols for downlink, 1 OFDM symbol for uplink, and 1 flexible OFDM symbol). Note 2 | #45 (6 OFDM symbols for downlink, 6 OFDM symbol for uplink, and 2 flexible OFDM symbol). Note 2 | #10 (13 OFDM symbol for uplink, and 1 flexible OFDM symbol). Note 2 |
| | 5G QoS Indicators (5QIs) | 5QI=80 (default priority level = 66, packet delay budget = 10 ms, packet error rate $10^{-6}$) | 5QI=4 (default priority level = 50, packet delay budget = 300 ms, packet error rate $10^{-6}$) | 5QI=81 (default priority level = 11, packet delay budget = 5 ms, packet error rate $10^{-5}$) |
| | Modulation and Coding Schemes (MCS) | (π/2) BPSK, QPSK, and 16/64/256 QAM | (π/2) BPSK, QPSK and 16/64 QAM | (π/2) BPSK, QPSK, and 16/64 QAM |
| Network management algorithms | e.g., Radio Resource Management (RRM); and Self-Organizing Network (SON) techniques | e.g., a dynamic scheduler for guaranteed throughput | e.g., a semi-persistent scheduler | e.g., a dynamic scheduler for guaranteed delay |
| Information derived by the RAN NSSMF | | | | |
| RUs covering the deployment area | | RUs of the city center | RUs of the entire city | RUs of the suburban area |
| Number of gNBs | | $N_1$ | $N_2$ | $N_3$ |
| Selected flavor in the gNB NSD | | Flavor #2 (eCPRI) | Flavor #3 (CPRI+eCPRI) | Flavor #1 (CPRI) |
| Subset of ILs from the selected flavor in the gNB NSD | | ILs per Region #3, i.e., [IL #(j+1), IL #k]. Note 3 | ILs per Region #1 (eCPRI) + Region #2 (CPRI) + Region #3 (eCPRI), i.e., [IL #(v+1), IL #w]. Note 3 | ILs per Region #2, i.e., [IL #(i+1), IL #j]. Note 3 |

**Table 3** Configuration parameters of each RAN NSST as well as the information derived by the RAN NSSMF. Note 1: Currently NR specifications do not provide any operation band supporting μ=4. Note 2: Flexible OFDM symbols might be used for both, downlink and uplink. Note 3: These levels match with those shown in Fig. 4.

Table 3 summarizes the configuration parameters of each RAN NSSTs as well as the information derived by the RAN NSSMF to instantiate the RAN slice subnets. Below, we discuss this information.

*eMBB*

For eMBB, the RAN NSST defines a numerology of μ=2 to fulfill the latency requirement of 10 ms. Any operation band supports this numerology. However, the adopted carriers should be used with the maximum available bandwidth to support the required high throughput.

Assuming the selection of TDD mode (common for the three use cases), the slot format #28 is set since it allocates the majority of slots for downlink traffic.

Concerning QoS classes, the RAN NSST per RAN determines the value #80 because it guarantees a latency lower than 10 ms.

The RAN NSST also specifies the utilization of the extended set of MCSs to provide the highest throughput values (i.e., those obtained from 256 QAM).

Considering the packet scheduling scheme as an example of network management algorithm, the RAN NSST selects a scheme that provides robust and adaptive data transmission. Particularly, the best option is a dynamic scheduler (as opposed to persistent scheduling) which also guarantees the throughput [15].

Finally, the RAN NSSMF selects the flavor #2 and the subset of ILs for Region #3 because they are adapted to the cell site density in a city center, and the fronthaul network of this region implements eCPRI for DU-RU interfaces.

### mMTC

For mMTC, the selected numerology is the lowest because the latency is not critical. Additionally, carriers' bandwidth should be the lowest possible, as the required throughput is low. Due to the small bandwidth, these carriers can only be allocated in the lower operation bands.

With respect to the 5QI, the RAN NSST selects the value #4, because it is the most latency-tolerant while the priority level is not too low.

Regarding the scheduling scheme, the semi-persistent scheduler is the best option since the traffic pattern of the sensors is deterministic, since the information is periodically exchanged with the network [15].

Finally, the RAN NSSMF selects the flavor #3 and the IL subset that considers a CU aggregating DUs serving three different regions over the entire city. Furthermore, this IL subset considers the implementation of eCPRI for the fronthaul networks of Region #1 and Region #3, and CPRI for the fronthaul network of Region #2.

### uRLLC

For uRLLC, the RAN NSST selects the highest numerology due to the stringent latency of 5 ms. This numerology forces the utilization of the highest operations bands. The slot format requires a larger amount of slots allocated in the uplink than in the downlink because vehicles continuously collect and send environment information to the remote drivers. Regarding 5QIs, only the #8 guarantees a latency below 5ms.

Finally, the RAN NSSMF selects the flavor #1 and the subset of ILs per Region #2 because they are adapted to the cell site density of the suburban area, and the fronthaul network of this region implements CPRI for DU-RU interfaces.

## Conclusions

RAN slicing enables the provision of different service types over a common wireless network infrastructure. Leveraging the NFV benefits, the CU and DUs of the gNBs for a RAN slice subnet could be customized and adapted to its requirements. Although the RAN NSST considers the gNB functionalities, the 3GPP has not identify which parameters and how they must be customized to provide a RAN slice subnet its expected behavior. Additionally, the RAN NSST neglects the resource requirements for the virtualized deployments of CUs and DUs over a geographical region with fluctuating spatial and temporal traffic demands. With the aim of enabling the customization and deployment of the gNBs of a RAN slice subnet, we have proposed a description model that harmonizes the 3GPP and ETSI-NFV viewpoints for RAN slicing. The proposed solution benefits from the reusability provided by NFV descriptors to define the underlying resources of the CU and DUs of the gNBs for several RAN slice subnets. To customize the behavior of each RAN slice subnet, we have identified the most representative radio parameters to configure their gNBs. Finally, to facilitate the comprehension of the proposal, an example composed of three RAN slice subnets for eMBB, mMTC and uRLCC scenarios has been provided.

## Acknowledgements

This work is partially supported by the Spanish Ministry of Economy and Competitiveness, the European Regional Development Fund (Project TEC2016-76795-C6-4-R), the Spanish Ministry of Education, Culture and Sport (FPU Grant 17/01844), and the University of Granada, Andalusian Regional Government and European Social Fund under Youth Employment Program.

## Author Information

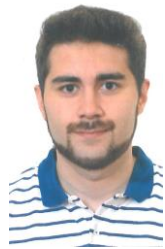
***Oscar Adamuz-Hinojosa*** *(oadamuz@ugr.es) received his B.Sc. and M.Sc. in Telecommunications Engineering from the University of Granada (Spain) in 2015 and 2017, respectively. He was granted a Ph.D. fellowship by the Education Spanish Ministry on September 2018. He is currently working toward his Ph.D. degree with the Department of Signal Theory, Telematics and Communication of the University of Granada. His research interests are focused on SDN, NFV and network slicing in 5G systems.*

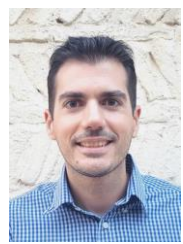
***Pablo Munoz*** *(pabloml@ugr.es) is Assistant Professor at the Dept. of Signal Theory, Telematics, and Communications of the University of Granada in Granada, Spain. He received the M.Sc. and Ph.D. degrees in telecommunication engineering from the University of Málaga, Málaga, Spain, in 2008 and 2013, respectively. He* has published more than 50 papers in peer-reviewed journals and conference proceedings and he is co-author of four international patents. His research interests include radio access network planning and management, application of artificial intelligence tools in radio resource management and virtualization of wireless networks.

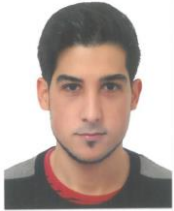

**Jose Ordonez-Lucena** *(joseantonio.ordonezlucena@telefonica.com) received his Master Eng. Degree in Telecommunications Engineering in 2017 from the University of Granada, Spain. He joined Telefonica I+D as a Core & Platforms Technology Analyst, within the GCTIO Unit. He is currently involved in technology exploration and innovation activities applicable to next-generation networks. Before joining Telefonica, he spent some years in the academic sector, dedicated to research on 5G mobile network architectures and network softwarization technologies, including SDN, NFV, and network slicing. From 2017, he is also pursuing a PhD in Telecommunications Engineering at the University of Granada.*

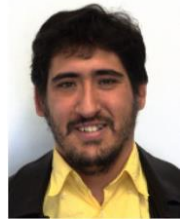

**Juan J. Ramos-Munoz** *(jjramos@ugr.es) received in 2001 his M.Sc. in Computer Sciences degree (2001) and Doctorate degree (2009) from the University of Granada, Spain. He is a Lecturer at the Department of Signals Theory, Telematics and Communications of the UGR. He is also member of the Wireless and Multimedia Networking Lab. His research interests are focused on real-time multimedia streaming, Quality of Experience assessment, network virtualization and network slicing for 5G.*

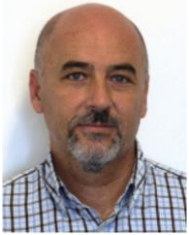

**Juan M. Lopez-Soler** *(juanma@ugr.es) is a professor in the Department of Signal Theory, Telematics and Communications, University of Granada. In 1991-1992 he joined the Institute for Systems Research at the University of Maryland. He is the head of the WiMuNet Lab at the University of Granada. He has participated in 24 research projects, has advised five Ph.D. theses, and has published more than 70 journal/conference papers. His research interests include middleware, multimedia communications, and 5G networking.*

## References

[1] S. E. Elayuobi *et al*., "5G RAN Slicing for Verticals: Enablers and Challenges," *IEEE Commun. Mag*., vol. 57, pp. 28-34, Jan. 2019.

[2] 3GPP TS 28.533, "Management and Orchestration; Architecture Framework (Release 15)," V15.5.0, Mar. 2019.

[3] ETSI GR NFV-EVE 012, "Network Functions Virtualization (NFV) Release 3; Evolution and Ecosystem; Report on Network Slicing Support with ETSI NFV Architecture Framework," Dec. 2017.

[4] R. Ferrus *et al.*, "On 5G Radio Access Network Slicing: Radio Interface Protocol Features and Configuration," *IEEE Commun. Mag.*, vol. 56, no. 5, pp. 184-192, May 2018.

[5] L. M. P. Larsen, A. Checko and H. L. Christiansen, "A Survey of the Functional Splits Proposed for 5G Mobile Crosshaul Networks," *IEEE Commun. Surveys Tuts.*, vol. 21, no. 1, pp. 146-172, Firstquarter 2019.

[6] C. Chang et al., "Slice Orchestration for Multi-Service Disaggregated Ultra-Dense RANs," *IEEE Commun. Mag.*, vol. 56, no. 8, pp. 70-77, August 2018.

[7] ITU-T, "GSTR-TN5G: Transport network support of IMT-2020/5G," Feb. 2018.

[8] 3GPP TS 23.501, "System Architecture for the 5G System; Stage 2 (Release 15)," V15.4.0, Dec. 2018.

[9] O. Adamuz-Hinojosa *et al.*, "Automated Network Service Scaling in NFV: Concepts, Mechanisms and Scaling Workflow", *IEEE Commun. Mag.*, vol. 56, no. 7, pp. 162-169, July 2018.

[10] 3GPP TS 38.211, "NR; Physical channels and modulation (Release 15)," V15.4.0, Dec. 2018.

[11] 3GPP TS 38.101-1, "NR; User Equipment (UE) radio transmission and reception; Part 1: Range 1 Standalone (Release 15)," V15.4.0, Dec. 2018.

[12] A. Younis, T. X. Tran and D. Pompili, "Bandwidth and Energy-Aware Resource Allocation for Cloud Radio Access Networks," in IEEE Transactions on Wireless Communications, vol. 17, no. 10, pp. 6487-6500, Oct. 2018.

[13] Next Generation Mobile Networks (NGMN), "NGMN 5G White Paper", [Online]. Available: https://www.ngmn.org/fileadmin/ngmn/content/images/news/ngmn_news/NGMN_5G_White_Paper_V1_0.pdf [Accessed: Jan. 2019].

[14] 3GPP TS 22.186, "Enhancement of 3GPP support for V2X scenarios; Stage 1 (Release 16)," V16.1.0, Dec. 2018.

[15] F. Capozzi, *et al.*, "Downlink Packet Scheduling in LTE Cellular Networks: Key Design Issues and a Survey," IEEE Commun. Surveys Tuts., vol. 15, no. 2, pp. 678-700, Second Quarter 2013.